\documentclass[english,aps,prl,amsmath,amssymb,twocolumn,letterpaper,showpacs]{revtex4-1}
\usepackage[T1]{fontenc}
\usepackage[latin9]{inputenc}
\usepackage{babel}
\usepackage{verbatim}
\usepackage{amsmath}
\usepackage{amssymb}
\usepackage{graphicx}
\usepackage{pdfpages}	% Packages needed
\usepackage{pgffor}		% to include supplement
\usepackage[unicode=true,pdfusetitle,
 bookmarks=true,bookmarksnumbered=false,bookmarksopen=false,
 breaklinks=false,pdfborder={0 0 1},backref=false,colorlinks=false]
 {hyperref}


\begin{document}

\title{Current noise from a magnetic moment in a helical edge}

\author{Jukka I. V\"ayrynen}

\author{Leonid I. Glazman}

\affiliation{Department of Physics, Yale University, New Haven, CT 06520, USA}

\date{\today}
\begin{abstract}
We calculate the two-terminal current noise generated by a magnetic
moment coupled to a helical edge of a two-dimensional topological
insulator. When the system is symmetric with respect to in-plane spin rotation,
the noise is dominated by the Nyquist component even in the presence of a voltage bias $V$. The corresponding noise spectrum $S(V,\omega)$ is determined by a modified fluctuation-dissipation theorem with the differential conductance $G(V,\omega)$ in place of the linear one. 
The  differential noise $\partial S/ \partial V$, commonly measured in experiments,  is strongly dependent
on frequency on a small scale $\tau_{K}^{-1}\ll T$ set by the Korringa
relaxation rate of the local moment. 
This is in stark contrast with the case of conventional mesoscopic conductors where $\partial S/ \partial V$ is frequency-independent and defined by the shot noise. 
In a helical edge, a violation of the spin-rotation symmetry leads to the shot noise, which becomes important only at a high bias. Uncharacteristically for a fermion system, this noise in the backscattered current is super-Poissonian. 
\end{abstract}
\maketitle

\textit{Introduction. } The edge conduction~\cite{2007Sci...318..766K,2009Sci...325..294R,2011PhRvL.107m6603K,2013NatMa..12..787N,Spanton04,Knez14,Li15}
in two-dimensional topological insulators~\cite{kane_z_2_2005,kane_quantum_2005,2006Sci...314.1757B,2008PhRvL.100w6601L}
(2D TI) such as HgTe and InAs/GaSb still poses unresolved questions. The lack of topological protection~\cite{2007Sci...318..766K,2009Sci...325..294R,2011PhRvL.107m6603K,suzuki_edge_2013,2014PhRvB..89l5305G,mueller_nonlocal_2015,PhysRevB.94.035301}
of the modes in both materials and the edge conductance in InAs/GaSb structures having similar values in the trivial and topological phases~\cite{1367-2630-18-8-083005,PhysRevLett.117.077701}, are not understood. The former question has created a plethora of theoretical activity in trying to identify the relevant time-reversal symmetric backscattering mechanism~\cite{Xu06,wu_helical_2006,2009PhRvL.102y6803M,tanaka_conductance_2011,lezmy_single_2012,2012PhRvL.108h6602B,schmidt_inelastic_2012,2013PhRvB..87p5440D,eriksson_spin-orbit_2013,2013PhRvL.110t6803C,Altshuler13,PhysRevB.90.075118,Yudson15,PhysRevB.92.205306,Chou15,PhysRevB.93.081301,PhysRevLett.116.086603}.
Charge puddles, ubiquitous in narrow-gap semiconductors and capable
of having a magnetic moment, are a plausible candidate to explain
the phenomena~\cite{vayrynen_helical_2013,vayrynen_resistance_2014}
and may have been observed in local measurements~\cite{2013PhRvX...3b1003K}.

A natural continuation to the measurements of average current is to measure correlations in it, \emph{i.e.}, noise. Theoretical studies
of noise in 2D TIs have focused on point-contacts and more complicated
geometries~\cite{PhysRevB.79.235321,Souquet12,PhysRevB.86.235121,PhysRevLett.110.246601,PhysRevLett.107.096602,PhysRevB.92.155421}.
The simpler case of two-terminal noise of a long single edge has been measured only recently~\cite{Tikhonov2015}. 
The observations were explained in Ref.~\cite{Aseev2016} where the authors found the conventional noise of a diffusive conductor at temperatures exceeding the charging energy of a near-edge puddle. 

\begin{figure}
\includegraphics[width=0.95\columnwidth]{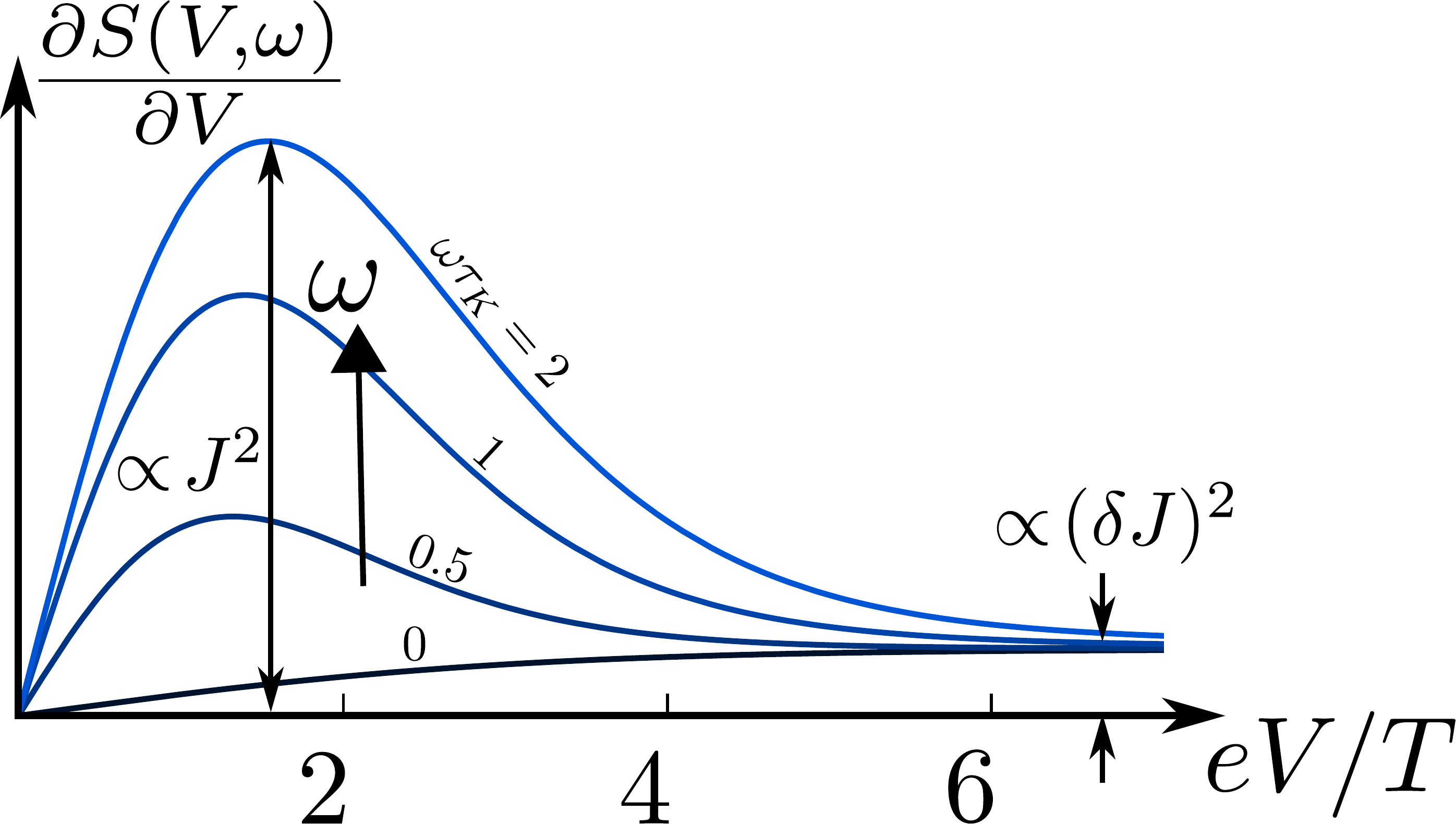}\caption{\label{fig:1}The low-frequency differential noise $\partial S/\partial V$, 
commonly measured by using bias modulation technique. At moderately low bias $eV\lesssim4T$ and $\omega\gtrsim\tau_{K}^{-1}$, 
the main contribution to $\partial S/\partial V$ comes  from the equilibrium noise, Eq.~(\ref{eq:diffnoise}); the shot noise is smaller by a factor $(\delta J/J)^{2}$. At higher bias,  the equilibrium noise contribution falls off exponentially with $eV/T$, and $\partial S/\partial V$ is dominated by the shot noise, Eq.~(\ref{eq:shotnoise}). Here $\delta J\! =\! 0.5\,J$, and the spin-flip rate $\tau_{K}^{-1}\!=\!0.025T$ at $V\! = \!0$.}
\end{figure}

At temperature $T$ below the charging energy, the puddle can develop a spin which gives rise to a new source of noise. 
In this paper we identify the signatures  in the noise caused by such a local magnetic moment. We calculate both equilibrium noise and shot noise. Due to a peculiar property of the local moment in a spin-orbit coupled system, the equilibrium noise is unconventional, and depends on the applied bias, see Fig.~(\ref{fig:1}). The most notable feature of the shot noise (which becomes important only at bias substantially exceeding $T$) is that it gives rise to  the Fano factor $F_{bs}>1$ in the backscattered current. The super-Poissonian nature of the noise is unusual for fermions and stems from the kinetics of the local spin flips which control the electron backscattering. Throughout this paper we use units $\hbar=k_{B}=1$. 

The coupling of the moment and the 2D TI helical edge can
be described by a weak exchange interaction. The full exchange tensor
consists of two parts, $J_{\text{full}}=J+\delta J$. The
larger piece $J$ is symmetric with respect to in-plane spin rotations~\cite{vayrynen_resistance_2014}
(hereafter we denote this symmetry as $U(1)_{s}$~\cite{maciejko_kondo_2012});
it does not lead to any dissipation of dc current~\cite{tanaka_conductance_2011}.
The exchange terms that break $U(1)_{s}$ are much smaller, $\delta J\ll J$,
but still important as they give rise to dc dissipation. 

The two parts of a generic  exchange interaction contribute
to the total current noise $S(V,\omega)$. Similarly, the noise
can be divided into two pieces, $S=S_{\text{eq}}+S_{\text{non-eq}}$. 

The larger $U(1)_{s}$ symmetric part $J$ of the exchange contributes only to $S_{\text{eq}}$.
(We call this component ``equilibrium noise''
since it satisfies a modified fluctuation-dissipation theorem.) 
Even though the exchange term $J$ does not modify the dc conductance of the edge, it gives rise to spin flips of the local moment. 
These local moment spin fluctuations are accompanied by backscattering events on the edge and therefore contribute to the  equilibrium current noise. 
The spin flip rate $\tau_K^{-1} \propto J^2 T$ defines the characteristic frequency of $S_{\text{eq}}(V,\omega)$. The application of a bias affects the distribution function of electrons in helical edge and results in its net spin density, which in turn polarizes the local moment. That polarization affects the noise power, making it bias-dependent. At high bias $eV \gg T$, the ``frozen'' local moment does not fluctuate and as a result the $J$-dependent part of $S_{\text{eq}}$ vanishes. 
The dependence on $V$ and the sensitivity to low frequencies, $\omega\ll T$,  make  $S_{\text{eq}}$ strikingly different from 
the conventional Johnson-Nyquist noise produced by electron elastic scattering~\cite{Blanter20001}. Since $S_{\text{eq}}$ depends on the bias voltage, it contributes to the differential noise $\partial S/ \partial V$. 

A common experimental technique to measure shot noise is the so-called bias modulation technique~\cite{PhysRevLett.75.3340,PhysRevLett.78.3370}
where the bias $V$ is modulated slowly in time leading in effect to a measurement of $\partial S/ \partial V$. 
In a conventional setup this eliminates the equilibrium (or thermal) noise, which is bias-independent. 
In contrast to this, in the case of a magnetic moment in a helical edge it is in fact the equilibrium noise $S_{\text{eq}}$ that dominates finite-frequency differential noise. 
%The above two properties of $S_{\text{eq}}$ have implications for 
%measuring shot noise using the bias modulation technique. 
%Our main finding is %the equation for
 We find the differential noise at $\omega \ll T$, 
\begin{equation}
\frac{\partial S(V,\omega)}{\partial V}  \! \approx  \frac{\partial S_{\text{eq}}(V,\omega)}{\partial V}  =   \frac{e^3}{2T\tau_{K}}\frac{(\omega\tau_{K})^{2}}{1+(\omega\tau_{K})^{2}} \frac{\tanh \frac{eV}{2T}}{\cosh^{2}\frac{eV}{2T}} \,.\label{eq:diffnoise}
\end{equation} %(-2T) \frac{d}{dV} \frac{1}{\cosh^{2}\frac{eV}{2T}} = 2e \frac{\tanh \frac{eV}{2T}}{\cosh^{2}\frac{eV}{2T}}
The second factor demonstrates the above-mentioned sensitivity of $S_{\textrm{eq}}$ to low frequencies ${\omega \sim \tau_{K}^{-1}}$, 
see also Fig.~\ref{fig:1}. 

\begin{figure}
\includegraphics[width=0.95\columnwidth]{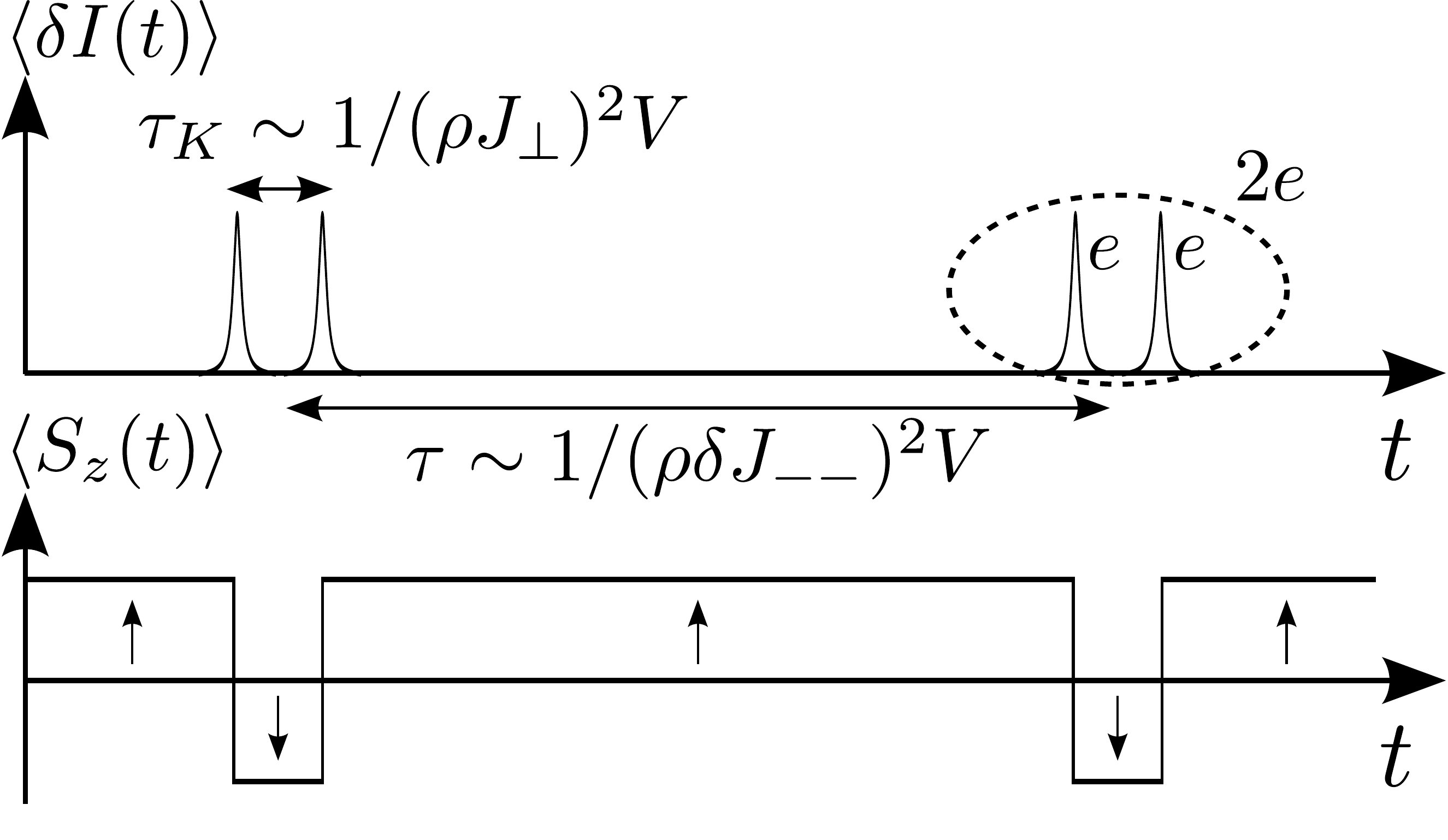}\caption{\label{fig:3}
Super-Poissonian shot noise  in the backscattered current $\delta I$. 
At large bias $eV \gg T$, the local spin spends most of its time in the quasiequilibrium state $\uparrow$, $\langle S_z \rangle = 1/2$. 
At random times, on average once per time $\tau \propto 1/ J_{--}^{2}$, the 
exchange term $\delta J_{--} S_- s_- +h.c.$ backscatters one electron and flips the localized spin ($\uparrow \to \downarrow$ in figure). After a short time $\tau_K \ll \tau$, the local moment relaxes to its quasiequilibrium state  ($\downarrow \to \uparrow $). In this process, another electron is scattered back.
Since 
$\tau \gg \tau_K$, the time-averaged backscattered current is $2e/\tau$ and the effective backscattered charge is $2e$ when probed at frequencies $\omega \ll 1/\tau_{K}$. 
}
\end{figure}

In the limit of high bias voltage, $eV\gg T$, the right-hand-side of Eq.~(\ref{eq:diffnoise}) vanishes. 
The differential noise is then dominated by the non-equilibrium component $S_{\text{non-eq}}$, which 
arises from the combination of the $U(1)_{s}$ breaking interaction $\delta J$ and finite bias voltage.
In the above limit, it % $S_{\text{non-eq}}$
becomes conventional shot noise
%$S_{\text{non-eq}} = e F_{bs} V \delta G_{\delta J}$, where  $\delta G_{\delta J} \propto \delta J^2$ is the reduction of ideal dc conductance. 
$S_{\text{non-eq}} = e F_{bs} \langle \delta I \rangle $, where  $\langle \delta I \rangle \propto \delta J^2 G_0 V $ is the  backscattering current (we denote $G_0 = e^2/h$ the edge ideal conductance). 
The backscattering current arises from rare random backscatterings of one electron which happen  with or without a local moment spin flip. 
%As we show in more detail below, 
In the former case, the subsequent  backscattering event will happen fast, in time $\tau_K$, and is correlated with the first one, see Fig.~\ref{fig:3}. 
This correlation results in super-Poissonian noise as demonstrated by the backscattering Fano factor $F_{bs} >1$. 
%Similarly, $F_{bs}$ will depend on the frequency on the scale $1/\tau_K$. 
%The \textit{total} current Fano factor $F=S_{\text{non-eq}} / e G_0 V$  is small, $F \propto \delta J^2 \ll 1$, as expected for  a nearly ballistic edge.   

Next, we move to a more detailed discussion of noise caused by a local magnetic moment on a helical edge. We start from the equilibrium noise and derive Eq.~(\ref{eq:diffnoise}).  

\textit{Equilibrium noise. } 
We start by considering the $U(1)_s$ symmetric system. 
The exchange Hamiltonian is then  $H_{\text{ex}}=\frac{1}{2}J_{\perp}(S_{-}s_{+}+S_{+}s_{-})+J_{z}S_{z}s_{z}$. 
Here $\mathbf{S}$ is the spin-operator of the localized moment, and  $\mathbf{s}=\frac{1}{2}\sum_{\alpha\beta}\psi_{\alpha}^{\dagger}\boldsymbol{\sigma}_{\alpha\beta} \psi_{\beta}$ is the spin density  of the helical edge electrons. 
We denote  $S_{\pm} = S_{x}\pm i S_{y}$. 
The Hamiltonian conserves the total spin 
$S_{z}^{\text{tot}}= S_z + \int dx s_z (x)$. 
In presence of an applied dc bias voltage $V$, the system is described by a generalized equilibrium Gibbs distribution  where the bias couples to the conserved quantity $S_{z}^{\text{tot}}$. 
Because of  spin-momentum locking, $\int dx s_z (x)$ is equal to the difference between the total number of right and left movers on the edge. 
The equilibrium density matrix 
\begin{equation}
\varrho\sim e^{-\beta\left(H-eVS_{z}^{tot}\right)}
\label{eq:rho}
\end{equation}
thus correctly assigns a chemical potential difference $eV$ between left and right movers. 
Having an equilibrium state, one may use a generalized fluctuation-dissipation theorem (FDT): 
\begin{equation}
S_{\text{eq}}(V,\omega)=\text{Re}G(V,\omega)\,\omega\coth\frac{\omega}{2T}\,.\label{eq:Sequilibrium}
\end{equation}
 Here $G(V,\omega)$ is the differential ac conductance of the helical edge defined
as the linear current response to a weak ac modulation $V_{1}e^{-i\omega t}$ on top of dc bias $V$. 
To calculate the noise $S_{\text{eq}}(\omega)$, it is thus sufficient to evaluate $G(V,\omega)$, outlined next. 

The current operator on the edge can be split into two %~\footnote{See Supplemental Material for details.  The supplement includes reference to Ref.~\onlinecite{2012PhRvB..86c5112L}. }
parts~\cite{Note1,KaneQH}, $I(t)=I_{in}(t)-\delta I(t)$, where  $I_{in}$ is the current incoming from the leads. 
The backscattering current $\delta I$ is given by the rate of change of the total number of right movers on the edge. 
By using the conservation of $S_{z}^{\text{tot}}$ and the total number of electrons~\cite{Note2}, %~\footnote{We consider low frequencies $\omega\ll v_{0}/L$ so that retardation effects can be neglected.},
 we can write the operator as $\delta I(t)=e\partial_{t} S_z$~\cite{tanaka_conductance_2011}. 
By employing the Kubo formula, this last form allows us to write the ac conductance as 
\begin{equation}
\text{Re} G(V,\omega)=G_0 - e^{2}\omega\chi''(V,\omega) \,,
\label{eq:G-chi}
\end{equation} where 
the second term $\chi''$ is the dissipative (imaginary) part of the local moment longitudinal susceptibility~\cite{Note3}. %\footnote{The dc bias voltage $V$ which polarizes the local moment has a role similar to an external magnetic field~\cite{PhysRevB.72.205125}.}. 

The local moment susceptibility  can be calculated by using the Bloch equation for  $\langle \mathbf{S}\rangle $~\cite{bloch46,Note1}. We find for $\omega\ll T$, 
\begin{equation}
 \chi''(V,\omega) = \frac{1}{4T\cosh^{2}\frac{eV}{2T}} \frac{\omega \tau_{K}}{1+(\omega\tau_{K})^{2}}\,. \label{eq:chi}
\end{equation}
Here 
 $\tau_{K}^{-1}=\pi T(\rho J_{\perp})^{2} \frac{eV}{2T} \coth\frac{eV}{2T}$ 
is the Korringa relaxation rate of the local moment, with $\rho$ being the density of states of the edge per spin per length (the low-bias limit of $\tau_K^{-1}$ was already used in the Introduction). 
% already mentioned in the introduction where the result for low bias $eV \ll T$ was used. 

%4. Mention Bloch equations for finding \chi(omega), give the result for Im\chi, explain \tau_K and its dependence on V (refer to Intro)
%5. Result for G(\omega,V), consequence for noise
Using Eq.~(\ref{eq:chi}) for $\chi''$ and its relation to $ \text{Re}\, G$, we find the ac conductance (valid at $\omega\ll T$)
\begin{equation}
\text{Re}G(V,\omega)=G_{0}-\frac{e^2}{4T\tau_{K}} \frac{1}{\cosh^{2}\frac{eV}{2T}} \frac{(\omega\tau_{K})^{2}}{1+(\omega\tau_{K})^{2}}\,,\label{eq:ReG}
\end{equation}
which generalizes the $V=0$ result of Ref.~\cite{tanaka_conductance_2011}
to arbitrary $V$. 
Substitution of Eq.~(\ref{eq:ReG}) into the FDT result, Eq.~(\ref{eq:Sequilibrium}), and subsequent differentiation yields  Eq.~(\ref{eq:diffnoise}). % after differentiation. 

%The ac conductance, Eq.~(\ref{eq:ReG}), rapidly drops once $\omega$ exceeds $\tau_{K}^{-1}$. 
%The noise,
%Eq.~(\ref{eq:Sequilibrium}), has a competing second factor which
%grows on the scale defined by $T$. Since at low bias $T\gg\tau_{K}^{-1}$ [see Eq.~(\ref{eq:korringa})], the equilibrium noise 
%will have a minimum in the interval $1/\tau_{K}<\omega<T$, and is
%strongly dependent on $\omega$ on a small scale $\tau_{K}^{-1}\ll T$~\cite{Note1}. 

At high bias voltage, $eV\gg T$, the noise %loses its sensitivity
%to low frequencies $\omega\sim\tau_{K}^{-1}$: the minimum moves to
%$\omega=0$, and the noise 
contribution from a $U(1)_{s}$ symmetric
part of exchange disappears, see Eqs.~(\ref{eq:ReG}) and (\ref{eq:Sequilibrium}). % Eqs.~(\ref{eq:omega_min}),~(\ref{eq:ReG})
%and Fig.~\ref{fig:1} inset.
This effect %of high bias
 can be understood
as follows. In the $U(1)_{s}$ symmetric case helical electron backscattering
is always associated with the local moment spin flip. At large dc
bias the moment is strongly polarized and has therefore exponentially small
susceptibility to spin flips caused by the time-varying bias,  see Eq.~(\ref{eq:chi}). This limitation does not concern $U(1)_{s}$ symmetry-breaking perturbations
which, for instance, enable backscattering of helical electrons without a flip of the  local moment. Such perturbations reduce the ideal zero bias conductance $G_0$ 
in Eq.~(\ref{eq:ReG}) by an amount $\sim (\rho\delta J)^{2}$~\cite{vayrynen_resistance_2014,vayrynen16a}, without affecting Eq.~(\ref{eq:diffnoise}). More importantly, the very same perturbation leads to the shot noise, $S_{\text{non-eq}}\propto(\delta J)^{2}V$, dominant in $S(V,\omega)$ at high bias.

\textit{Non-equilibrium noise; shot noise and Fano factor.}
At finite bias $V$, the perturbation $\delta J$ drives the density matrix out of the equilibrium state Eq.~(\ref{eq:rho}). We focus on the limit of high bias voltage, $eV \gg T$, and evaluate the backscattered current noise within the perturbation theory in $\delta J$. 
%At finite bias $V$, the perturbations $\delta J$ drive the system out of equilibrium. In this section, we  discuss the resulting non-equilibrium noise $S_{\text{non-eq}}$. We focus on the limit of high bias voltage, $eV \gg T$, where $S_{\text{non-eq}}$ dominates the total noise.  Unlike in the case of $S_{\text{eq}}$ where FDT could be used, the calculation of $S_{\text{non-eq}}$ does not simply reduce to calculation of ac conductance. Instead, we calculate (the Fourier transform of) 
The operator of backscattered current is 
 $\delta I(t)=e\partial_{t}S_{z}-e\partial_{t}S_{z}^{\text{tot}}$, 
which generalizes the earlier form $\delta I(t)=e\partial_{t}S_{z}$ to the case of broken $U(1)_{s}$  symmetry. The full symmetrized current noise is  ${S(t) \!= \! \text{Re}\left\langle I(t+t')\negthinspace:\negthinspace I(t')\negthinspace:\right\rangle}$, where $:\!\! I(t)\!\! :=I(t)-\left\langle I\right\rangle $ and $I=I_{in}-\delta I$. 

At high bias, $eV\gg T$ the full noise is dominated by the shot noise, obtained from the $\delta I$--$\delta I$ correlation function. 
The term $e \partial_{t}S_{z}$ in the operator $\delta I$ does not contribute to the zero-frequency noise~\cite{Note1}, and we find~\cite{Note4}, %\footnote{At low frequencies, $S_{\text{non-eq}}$ varies on the scale $1/\tau_{K}$, see Ref.~\cite{nagaev06}. At high frequency,  $\omega \sim eV$, the shot noise has singularities at $\omega = \pm n eV$~\cite{Basset12}.}
\begin{equation}
S(V,0)\!\approx\! S_{\text{non-eq}}(V,0)\!=\!e\sum_{n=1,2}n\langle\delta I^{(n)} \rangle
\,.\label{eq:shotnoise}
\end{equation}
 Here $\langle\delta I^{(n)}\rangle$ is the contribution to backscattered current from the terms of the exchange interaction which change $S_{z}^{\text{tot}}$ by $n\hbar$. The origin of the prefactor $n$ is explained below (see also Fig.~\ref{fig:3} which illustrates the terms with $n=2$). 

The resulting Fano factor, $F =S(V,0)/e\langle I\rangle$, to lowest order in
$\delta J$ is $F\sim(\rho\delta J)^{2}\ll1$ as expected for weak
backscattering~\cite{Blanter20001}. (Small Fano factors, $F\lesssim0.26$,
were found in the experiment~\cite{Tikhonov2015} in HgTe and in the theoretical paper Ref.~\cite{Aseev2016}.) The backscattering Fano factor defined as $F_{bs}={S(V,0)}/{e\langle\delta I\rangle}$ lies between
$1$ and $2$: 
\begin{equation}
F_{bs} =1+\frac{|\delta J_{--}|^{2}}{|\delta J_{--}|^{2}+\frac{1}{2}\sum_{i=x,y}[\delta J_{zi}+\delta J_{iz}]^{2}}\,.\label{eq:estar}
\end{equation}
Here $\delta J_{--}=\delta J_{xx}-\delta J_{yy}-i{(\delta J_{xy}+\delta J_{yx})}$
corresponds to the exchange component that changes $S_{z}^{\text{tot}}$ by
$2\hbar$ (while $\delta J_{zi}$ and $\delta J_{iz}$ change it by $\hbar$)~\cite{vayrynen16a}. 

It is tempting to interpret $F_{bs}$ as the effective charge (in units of $e$) of the backscattered particles~\cite{PhysRevLett.97.086601}. 
In our considered system this interpretation must be used with caution: we find a super-Poissonian value $F_{bs}>1$, Eq.~(\ref{eq:estar}), even though the exchange Hamiltonian only backscatters a single electron at a time.
The reason behind  $F_{bs}>1$ is correlated backscattering that employs the exchange terms $\delta J_{--}S_- s_- +h.c.$ and $\frac{1}{2} J_{\perp}S_+ s_- +h.c.$ sequentially, as demonstrated in Fig.~\ref{fig:3}. 
The corresponding times $\tau\sim 1/(\rho \delta J_{--})^2V$ and $\tau_K\sim 1/(\rho J_\perp)^2V$ are vastly different, $\tau_K\ll\tau$, because $\delta J/J\ll1$. As the result, the spin-flip event caused by $ \delta J_{--}$ which drives system out of the equilibrium, is quickly followed by a flip restoring the dominant spin direction, see the lower panel of Fig~\ref{fig:3}. Each of the flips sends one electron back. Therefore, electrons are backscattered in pairs. The characteristic time $\tau_K$ separates the electrons in one pair, while $\tau$ is the time between pairs (see the upper panel of Fig~\ref{fig:3}). These paired events are uncorrelated with each other, forming a Poissonian process with the average backscattered  current $2e /\tau$ and the corresponding Fano factor 2. Inclusion of the backscattering processes without a flip of the local moment, {\it e.g.}, due to term $S_z s_x$ in the exchange interaction, yields $1< F_{bs} < 2$, Eq.~(\ref{eq:estar}). 

The exchange interaction $J$ of a local moment with itinerant electrons leads to the Kondo effect~\cite{1964PThPh..32...37K} which, in the absence of external perturbations, results in screening of the local spin at low temperatures $T\lesssim T_K$. The Kondo effect is a crossover rather than a phase transition occurring at some temperature. The screening builds up gradually upon reducing $T$ towards $T_K$, and at high temperatures may be accounted for by means of the renormalization group (RG) technique. Within RG, the ``bare'' exchange constants are replaced by running ones, $\rho J(D)=1/\ln(D/T_K)$ and $\rho\delta J(D)=\rho \delta J(D_0) \ln (D/T_K) / \ln (D_0/T_K)$ which are functions of energy scale $D$. (The bare constants depend on the microscopic origin of the exchange interaction; the corresponding ``bare'' Hamiltonian is defined in the band of width $D_0>D$.) As long as $D_0\gtrsim \mathrm{max} (T,eV) \gtrsim T_K$, we may account for the Kondo effect in noise by replacing $J,\delta J\to J(D), \delta J(D)$ with $D=\mathrm{max}(T,eV)$. Going to lower $T, eV$ requires a more careful consideration, discussed next. 

%So far, we have assumed that %$T$ or $eV$ is above the local moment Kondo temperature, so that 
%%the moment is not strongly screened, and
% the exchange couplings are small, $\rho J\ll 1$. 
%Upon lowering temperature and bias, this assumption will eventually become invalid due to the renormalization of the exchange couplings. 
%The  Kondo temperature $T_K$ is defined as the scale at which perturbation theory fails,  $\rho J (T_K)\sim 1$. 
%Below this temperature, the local moment becomes strongly coupled to the edge electrons. Next, we study the noise in this limit.

\textit{Noise at low temperatures $T<T_{K}$. }
Let us begin our discussion from the $U(1)_{s}$ symmetric case,
$\delta J=0$, in which the noise,  Eq.~(\ref{eq:Sequilibrium}), is determined by the spin susceptibility $\chi''(V,\omega)$, Eq.~(\ref{eq:G-chi}). 
At low frequencies and temperature ($\omega,\,T\ll T_K$) the susceptibility is known from the Shiba  relation~\cite{1975PThPh..54..967S} generalized to finite magnetic field~\cite{PhysRevB.72.205125},
 whose role here is played by the bias voltage. 
We find~\cite{Note1} $\chi'' (V,\omega) = c_1 \frac{\omega}{T_K^2} f(eV/T_K)$. 
% where the function $f(x)$ has asymptotes~\cite{hewson97} $f(x) \approx 1 - c_2 x^2$ for $x\ll 1$ and $f(x) \approx 1 /(W x \ln x)^2 $ for $x\gg 1 $. 
The differential noise at $\omega,\, T \ll T_K$  is then~\cite{Note5}, %~\footnote{The numerical factors are $c_1 =2\pi (W/4)^2$, $c_2 = 3 \sqrt{3} c_1 /4$ with $W\approx0.4$ being the Wilson number~\cite{hewson97}.}, 
\begin{align}
\frac{\partial S}{\partial V} \! &= \!  c_1 e^3 \frac{ \omega^3}{T_K^3}  \coth \frac{\omega}{2T}\left[-f'\left(\frac{eV}{T_K}\right)\right]\,, \\
f'(x) &= -\begin{cases}
2 c_2 x\,, & x\ll 1\,,\\
%\frac
{2}[{W^2} 
%\frac{1}
{x^3 \ln^2 x}]^{-1}  \,, & x \gg 1\,.
\end{cases}
\end{align}
The $\partial S/\partial V$ vs. $V$ dependence is qualitatively similar to the one in  Fig.~(\ref{fig:1}) with the replacement $T\to T_K$ on the horizontal axis. 

The equilibrium differential noise described above vanishes at ${\omega =0}$, as well as in the limit of high bias $eV\gg T_K$; in these regimes the noise is dominated by the non-equilibrium component coming from exchange interaction terms breaking the $U(1)_s$ symmetry. 
In the high-bias limit the discussion of the previous section, including Eq.~(\ref{eq:shotnoise}), remains valid. 
In the opposite regime, $eV \ll T_K$, 
 we can use   an effective local interaction Hamiltonian~\cite{1974JLTP...17...31N,schmidt_inelastic_2012,lezmy_single_2012} to describe the breaking of $U(1)_s$ symmetry. 
The most important irrelevant perturbation backscatters one particle inelastically. The low-frequency shot noise,  Eq.~(\ref{eq:shotnoise}), is then given by the $n=1$ term: at $T,\, eV \ll T_K$ we have  $\langle\delta I^{(1)} \rangle \gg \langle\delta I^{(2)} \rangle$~\cite{lezmy_single_2012}. 
As a result, the backscattering Fano factor crosses over from $F_{bs}>1$, Eq.~(\ref{eq:estar}), to $F_{bs}=1$ as $V$ is decreased below $T_K$. 
This $V$-dependence of $F_{bs}$ is not visible in Eq.~(\ref{eq:estar}) in the leading-order RG, since the corresponding running couplings $\delta J$ have the same $V$-dependence. On the contrary, the Fano factor $F$ of the total current noise shows reduction from $F_\infty\sim[\rho\delta J(D_0)]^2$ to even smaller values $F(V)\sim F_\infty[\ln(eV/T_K) / \ln (D_0/T_K)]^2$ when the bias is reduced from $\sim D_0$ to $eV \gtrsim T_K$. Matching this RG result with the known~\cite{lezmy_single_2012} power-law-dependence in the low-bias regime, we find $F(V) \sim F_\infty [\ln (D_0/T_K) ]^{-2} (eV/T_K)^4$ at $eV \lesssim  T_K$.

\textit{Discussion. }
Measurement of noise provides ways to distinguish between different sources of backscattering on the helical edge. 
In particular, it may tell the difference between the effects of a local magnetic moment and strong electron-electron interactions on the edge; both explanations are consistent with  recent conductance measurements~\cite{Li15,vayrynen16a}. The  zero-frequency shot noise provides a clear difference between the two backscattering mechanisms. 
In both cases there is a crossover in the bias-voltage dependence of the Fano factor $F(V)$. In the case of strong electron-electron interactions, $F(V)$ is a  monotonically-decreasing function with a fairly weak bias dependence: $1-F \propto V^{\alpha}$ below some crossover bias value ($V\ll V^*$), and $F\propto V^{-\alpha}$ above the crossover, with $0< \alpha \ll 1$.
%the cross over is between the limits of weak ($V\gg V^*$) and strong backscattering ($V\ll V^*$), separated by a crossover scale $V^*$. In the former limit of high bias, the Fano factor is small and has a weak power law $V^{-\alpha}$ with $0< \alpha \ll 1$. Below the crossover scale, the edge becomes insulating and Fano factor approaches 1 as $1-F \sim V^{\alpha}$ with the same small exponent. 
In the case of a local magnetic moment, the Fano factor $F$ monotonically increases with $V$, exhibiting a highly asymmetric crossover around $eV \sim T_K$. Below the crossover scale, we have a strong power-law $F(V) \propto V^4$ (see end of the previous paragraph) while above it the $V$-dependence is weak, $F(V) %\propto (\rho \delta J)^2
 \propto \ln^2 (eV/T_K) $. (Weak electron-electron interaction may change the logarithmic function here to a weak power-law~\cite{vayrynen16a}.) Moreover, the Fano factor remains small, $F\ll 1$, at any bias. 

Another clear distinction between magnetic moments and other backscattering mechanisms is in the finite-frequency noise power $S(V,\omega)$. In the case of a magnetic moment, the differential noise $\partial S/ \partial V$ at frequencies  $\omega \sim \tau_{K}^{-1}$ is dominated by the bias-dependent equilibrium  component,
%$\propto J^2$ 
see Fig.~(\ref{fig:1}). The characteristic time $\tau_K$ here is determined by the relaxation dynamics of the local moment. In the conventional case of a scatterer having no internal dynamics, $\partial S/ \partial V$ is produced by the  shot noise which is insensitive to frequencies all the way up to $\omega \sim eV$~\cite{PhysRevLett.78.3370}. 
In the case of magnetic moment, the shot noise contribution
is much smaller than the equilibrium one, as it requires a perturbation breaking the $U(1)_s$ symmetry. For a spin-carrying charge puddle, it is generated by spin non-conserving interaction in it; one can estimate the ratio of the corresponding exchange constant to the one respecting the $U(1)_s$ symmetry as $ \delta J /J \sim 1/g$
%$J / \delta J \sim g$ 
where $g\geq 1$ is the dimensionless conductance of the puddle~\cite{vayrynen_resistance_2014}. (We took $g =2$ in Fig.~\ref{fig:1} to illustrate that even a fairly small value for $g$ gives a sizable effect.)

Besides magnetic moments due to charge puddles, our theory is applicable to the case of magnetic doping, \textit{e.g.}, Mn impurities in InAs. 
Such impurities have an out-of-plane magnetic easy axis which does not violate the $U(1)_s$ symmetry~\cite{Wang2014}. 
Therefore we expect the equilibrium noise to remain the most  important noise component. 
Furthermore, the magnetic anisotropy energy is estimated to be  small, $\mathcal{K} \approx 0.08 \mathrm{K}$ in InAs host~\cite{PhysRevB.63.054418}; noise-producing spin flips are thus expected to occur even at fairly low temperatures.  
To estimate the spin-flip rate $\tau_K^{-1}$ we can use the value $\rho J (D_0)=0.03$~\cite{vayrynen16a} which gives the frequency scale $\tau_{K}^{-1}\approx 80$~MHz at $T=1$~K. 
This frequency is safely above $1/f$ noise; in measurements~\cite{Tikhonov2015} of the shot noise in a HgTe heterostructure
 even lower frequencies, $10-20$~MHz, were successfully used.
 
\begin{acknowledgments}
We thank Rui-Rui Du, Tingxin Li, and Ivana Petkovic for discussions. This work was supported by NSF DMR Grant No. 1603243.
\end{acknowledgments}

%\bibliographystyle{apsrev4-1}
%\bibliography{refs}

%

% The following merges the supplement into the main text


\section*{Supplementary material}

# Supplemental Material to "Current noise from a magnetic moment in a helical edge"

Jukka I. Väyrynen and Leonid I. Glazman
*Department of Physics, Yale University, New Haven, CT 06520, USA*

In this supplemental material we show in detail the technical derivations omitted from the main text.

## SM1. THE CURRENT OPERATOR

In the main text, we divided the current operator into incoming and backscattered components: $I(t) = I_{in}(t) - \delta I(t)$. In this section we explain the details of this separation.

The current operator $I(t)$ is defined as

$$I(t) = ev[n_R(x,t) - n_L(x,t)]_{x=0}, \qquad (S1)$$

where $n_{R(L)}$ is the density operator of right-moving (left-moving) electrons, and $v$ is their velocity. We evaluate the current operator at the position of the left lead at $x = 0$. The operator in Eq. (S1) does not contain any explicit dependence on the perturbations causing backscattering on the edge (only its time-evolution and average with the density matrix do). It is therefore useful to split the operator into two parts, one of which includes all the dependence on backscattering explicitly. We write $I(t) = I_{in}(t) - \delta I(t)$, where $I_{in}$ is the current incoming from the leads at $x = 0, L$ ($L$ is the length of the edge) and $\delta I$ is the backscattered current. The two contributions are

$$I_{in}(t) = ev[n_R(0,t) - n_L(L,t)], \quad \delta I(t) = -e\frac{d}{dt}\int dx s_z(x), \qquad (S2)$$

as can be derived from current and particle number conservation. Here $\int dx s_z(x) = \frac{1}{2}\int_0^L dx\,[n_R(x) - n_L(x)]$ commutes with the $s_z$-conserving part of the Hamiltonian. Upon presenting the operator of backscattering current as $-ei[H, \int dx s_z(x)]$, it is clear that it contains only contributions from perturbations that backscatter electrons. (We assume that all the backscattering happens in the interval $0 < x < L$.)

## SM2. DERIVATION OF EQS. (4)–(6) OF THE MAIN TEXT BY USING THE BLOCH EQUATION

In the $U(1)_s$ symmetric case the Hamiltonian conserves the total spin projection along $z$-axis; it commutes with

$$S_z^{\text{tot}} = \int dx s_z(x) + S_z. \qquad (S3)$$

As was mentioned in the main text, Eq. (2), the density matrix of the system has then the equilibrium form $\varrho \propto e^{-(H-eVS_z^{\text{tot}})/T}$ even in the presence of a bias voltage $V$. This equilibrium form allowed us to use a modified fluctuation-dissipation theorem to calculate the current noise, see Eq. (3) of the main text. In that equation

$$G(V,\omega) = \lim_{V_1 \to 0} \langle I \rangle / V_1 e^{-i\omega t} = G_0 - \lim_{V_1 \to 0} \langle \delta I \rangle / V_1 e^{-i\omega t}, \qquad (S4)$$

is the differential ac conductance, calculated as a response to weak ac modulation of the bias voltage $V_{ac}(t) = V + V_1 e^{-i\omega t}$. In Eq. (S4) we divided the current operator into the two parts, Eq. (S2), and wrote the trivial contribution $G_0 = e^2/h$ coming from the incoming current. The second contribution to Eq. (S4), coming from the backscattering current can be written as $\delta I = e\partial_t S_z$ [S1] by using Eq. (S3) with $dS_z^{\text{tot}}/dt = 0$.

To evaluate the average $\langle \delta I \rangle = e\langle \partial_t S_z \rangle$ at $\omega \ll T$, we use the Bloch equation [S2, S3]

$$\frac{d}{dt}\langle S_z(t) \rangle = -\tau_K^{-1}[\langle S_z(t) \rangle - S_z^f(t)], \qquad (S5)$$

where $\tau_K^{-1} = \pi T(\rho J_\perp)^2 \frac{eV}{2T}\coth\frac{eV}{2T}$ is the Korringa rate and $S_z^f(t) = \frac{1}{2}\tanh\frac{eV_{ac}(t)}{2T}$ is the instantaneous equilibrium polarization obtained from the distribution $\varrho$ [S4]. (To linear order, we can ignore the influence of $V_1$ on the Korringa rate.)

At long times $t \gg \tau_K$ we find to first order in $V_1$,

$$\langle S_z(t)\rangle = \frac{eV_1 e^{-i\omega t}}{4T} \frac{1+i\omega\tau_K}{1+(\omega\tau_K)^2}\frac{1}{\cosh^2\frac{eV}{2T}} + \frac{1}{2}\tanh\frac{eV}{2T}, \tag{S6}$$

and thus

$$\langle \delta I\rangle = e^2 \frac{V_1 e^{-i\omega t}}{4T\tau_K} \frac{-i\omega\tau_K(1+i\omega\tau_K)}{1+(\omega\tau_K)^2}\frac{1}{\cosh^2\frac{eV}{2T}}. \tag{S7}$$

Using Eq. (S7) in Eq. (S4) gives Eq. (6) of the main text. Note that by using Eqs. (S4), (S6) and (S7), we can relate $G(\omega) - G_0$ to the spin susceptibility

$$\chi(V,\omega) = \lim_{V_1\to 0}\left(\langle S_z(t)\rangle - \frac{1}{2}\tanh\frac{eV}{2T}\right)/eV_1 e^{-i\omega t} = \chi_0(V,T)\frac{1+i\omega\tau_K}{1+(\omega\tau_K)^2}, \tag{S8}$$

where we introduced the static susceptibility $\chi_0(V,T) = (4T\cosh^2\frac{eV}{2T})^{-1}$. Note that the bias voltage has a role similar to an external magnetic field [S5].

### SM3.  THE DISSIPATIVE PART OF SPIN SUSCEPTIBILITY AT $\omega, T \ll T_K$

To evaluate the imaginary part of the susceptibility $\chi''(V,\omega)$, we use the generalized Shiba relation [S5]

$$\chi''(V,\omega) = 2\pi\omega\left(\frac{d\langle S_z\rangle}{deV}\right)^2, \tag{S9}$$

where in our case the role of external static magnetic field is played by the bias $eV$. The polarization $\langle S_z\rangle$ in the scaling limit can be found in the high and low bias limits [S6],

$$\langle S_z\rangle = \begin{cases} \frac{1}{2}\left(1 - \frac{1}{2}\frac{1}{\ln eV/T_K} + \ldots\right), & eV \gg T_K, \\ \frac{W}{4}\frac{eV}{T_K}\left(1 - \frac{\sqrt{3}\pi W^2}{64}\left(\frac{eV}{T_K}\right)^2 + \ldots\right), & eV \ll T_K, \end{cases} \tag{S10}$$

where $W$ is the Wilson number. By calculating the susceptibility from Eq. (S9) with the help of Eq. (S10), we find the form

$$\chi''(V,\omega) = 2\pi \frac{W^2}{16}\frac{\omega}{T_K^2}\left\{\begin{matrix}[W x\ln x]^{-2}+\ldots, & x\gg 1,\\ 1-2\frac{3\sqrt{3}\pi W^2}{64}x^2+\ldots, & x\ll 1,\end{matrix}\right\}_{x=eV/T_K}, \tag{S11}$$

which gives Eq. (10) of the main text. In particular, the function inside the curly brackets is $f(x)$ of the main text.

### SM4.  THE TERM $\partial_t S_z$ DOES NOT CONTRIBUTE TO ZERO-FREQUENCY NOISE POWER

In this section, we show that the first term in the current operator, $\delta I = e\partial_t S_z - e\partial_t S_z^{\text{tot}}$, does not contribute to the zero-frequency noise power

$$S(V,0) = \int_{-\infty}^{\infty} dt\,\mathrm{Re}\,\langle I(t+t'){:}I(t'){:}\rangle. \tag{S12}$$

We will write $I = I_{in} - \delta I = A - e\partial_t S_z$ where $A = I_{in} + e\partial_t S_z^{\text{tot}}$. (The operators without time-argument are evaluated at $t=0$.) Upon inserting into Eq. (S12), we find 3 terms that involve the operator $\partial_t S_z$. The two terms involving the operator $A$ give equal contributions to Eq. (S12), as one can show by changing variables $t \to -t$ and using the properties of the real part of the correlation function. All of the 3 terms can then be cast in the form

$$\int_{-\infty}^{\infty} dt\,\mathrm{Re}\,\langle \partial_t S_z(t){:}B{:}\rangle = \mathrm{Re}\,\langle S_z(t){:}B{:}\rangle\big|_{-\infty}^{\infty} = 0, \quad \text{with } B = A \text{ or } B = -e\partial_t S_z. \tag{S13}$$

The last equality assumes that the system is in equilibrium in the infinite past and future; the perturbation (in our case the symmetry breaking exchange term $\delta J$) acts for a finite (but long) amount of time.



\end{document}